\documentclass[12pt]{article}
  \usepackage{amsthm,amsfonts}
  \usepackage{amsmath}
\newcommand{\bea}   {\begin{eqnarray}}
\newcommand{\eea}   {\end{eqnarray}}
\begin{document}
\renewcommand{\thefootnote}{\fnsymbol{footnote}}

\thispagestyle{empty}

\title{Wigner Oscillators, Twisted Hopf Algebras and Second Quantization}
\author{P. G. Castro\thanks{{\em e-mail: pgcastro@cbpf.br}}, B. Chakraborty\thanks{{\em e-mail: biswajit@bose.res.in}}
 ~and F.
Toppan\thanks{{\em e-mail: toppan@cbpf.br}}
\\ \\
{\it $~^\dagger$ S. N. Bose National Center for Basic Sciences,}\\{\it JD Block, Sector III, Salt-Lake, Kolkata-700098, India}
\\
{\it $~^{\ast\dagger\ddagger}$ CBPF, Rua Dr.}
{\it Xavier Sigaud 150,}
 \\ {\it cep 22290-180, Rio de Janeiro (RJ), Brazil}}
\maketitle
\begin{abstract}
By correctly identifying the role of central extension in the centrally extended Heisenberg algebra $h$, we show that it is indeed possible to construct a 
Hopf algebraic structure on the corresponding enveloping algebra $\mathcal{U}(h)$ and eventually deform it through Drinfeld twist. This Hopf algebraic structure and its deformed version $\mathcal{U}^{\mathcal F}(h)$ are shown to be induced from a more
``fundamental"  Hopf algebra obtained from the Schr\"odinger field/oscillator algebra and its deformed version, provided that the fields/oscillators are regarded as odd-elements of the super-algebra $osp(1|2n)$. We also discuss the possible implications in the context of quantum statistics.

\end{abstract}
\vfill

\rightline{CBPF-NF-009/08}

\newpage
\section{Introduction}
The initial activities in noncommutative (NC) quantum field theories \cite{szabo1}  were plagued by the problem of  violation of the Poincar\'e symmetry.
In its simplest version, one introduces the matrix-valued noncommutative parameter $\Theta=\{\theta^{\mu \nu}\}$ through the commutation relation
$[{\hat x}^{\mu }, {\hat x}^{\nu }]=i\theta^{\mu \nu}$. At the level of this commutation relation the Lorentz covariance clearly implies that this antisymmetric
object $\theta^{\mu \nu}$ should transform as a second rank contravariant tensor. The presence of such a constant tensor-valued parameter,
which acts like a constant background field,
however, generates a certain torque on the system, thereby modifying the criterion for Lorentz invariance \cite{ban} in the form of
yielding non-vanishing 4-divergence of the angular momentum tensor. With these modified criteria, it is indeed possible to verify the
Lorentz invariance properties of the various actions in NC QFT's. For that one usually writes effective commutative actions using Seiberg-Witten
map. However, here one has to terminate the series to a certain order of the NC parameter. Besides, the physical equivalence of these
two versions is not clear, as it has been shown explicitly earlier, in a simple quantum mechanical context, that the Seiberg-Witten
flow in the non-commutative parameter is not spectrum preserving in presence of interactions \cite{scholtz}. Furthermore, one can show, by considering
the vacuum expectation value of the above mentioned commutator $[{\hat x}^{\mu }, {\hat x}^{\nu }]$, that the
presence of such a transforming $\Theta $ gives rise to spontaneous violation of Lorentz symmetry. On the other hand, one would like to hold all the components of the
$\Theta $ matrix fixed in all the Lorentz frames, so that these can be elevated to the status of some new fundamental constants of Nature
like $\hbar , c, G$ etc. \cite{aschieri1}, as required by certain Gedanken experiment \cite{doplicher, aschieri} trying to probe spacetime structures at Planck level.
However, in this case the Lorentz symmetry is violated explicitly. The only way which can reconcile constant and non-transforming
$\theta^{\mu \nu}$ with Poincar\'e symmetry is by a twisted implementation of this group in a Hopf algebraic setting by using certain abelian twist
{\it \`a la} Drinfeld \cite{chaichian}. Since then there has been an upsurge of interest in the study of various implications of non-commutativity and its experimental
consequences within this framework. This was extended later to the case of Galilean symmetry \cite{chakraborty}-relevant for the nonrelativistic systems- and
it was shown that Pauli principle can be violated by the so called ``twisted fermions". This indicates that the twist can have a non-trivial effect on
quantum statistics as well. However, most of the studies were confined to the relativistic
QFT. Besides, even in  \cite{chakraborty}, the Schr\"odinger field was considered as $c \rightarrow \infty$ limit of the corresponding relativistic field (reminiscent
of Wigner-\.In\"on\"u group contraction) and an explicit Hopf algebraic deformation of the Schr\"odinger fields/oscillators were not considered and just the
twisted anti-commutation relations were exploited. This is an important question to be addressed in order to analyse noncommutative quantum
mechanical systems consistently, where one has to deal with the Heisenberg algebra $h$ involving the composite position and momentum operators,
which are known to be given by certain integrated objects of bilinears of Schr\"odinger field operators. A natural follow-up of this question is whether
the above mentioned deformation at the level of Schr\"odinger oscillators can induce appropriate deformations at the level of the position and
momentum operators as well. For this, one has to first understand whether one can, at all, construct a Hopf algebra structure on Heisenberg algebra
or, for that matter, any Lie algebra with central extension. In fact, this issue was addressed earlier in the literature \cite{palev}, where it was shown that one
cannot construct Hopf algebra structure on Heisenberg algebra. However, in this analysis the central extension was regarded as a mere multiple
of the identity belonging to the corresponding Lie group, so that the central extension has the same co-product structure as that of the identity element.

The purpose of the present paper is to first revisit the above mentioned issues. Particularly, we find that it is indeed possible to construct a Hopf algebra structure on Heisenberg algebra or, more precisely, on the corresponding enveloping algebra $\mathcal{U}(h)$, by carefully re-interpreting the role of central
extension. This also enables one to deform it by using Drinfeld twist. In particular, we shall be using the abelian twist appropriate for Moyal star product.
With this, we reproduce most of the existing results in the literature. At the next stage, we show how the Hopf algebraic structure and its deformed
version can be similarly constructed out of the Schr\"odinger oscillator algebra, which in turn can induce an appropriate Hopf algebraic structure on the composite
Heisenberg operators, where we observe the important roles played by the super-algebras. We also note the important consequences of these
constructions in quantum statistics.

The paper is organized as follows. In section \textbf{2}, we provide a brief review of Hopf algebras, collecting all the important formulae to be used in the
subsequent sections. We then apply this formalism in section \textbf{3} to construct the Hopf algebra structure on $\mathcal{U}(h)$, involving bosonic
variables only. The corresponding fermionic case is taken up in section \textbf{4}, which is used subsequently in the context of super-algebra. In the following
section \textbf{5} we introduce the second quantized Schr\"odinger fields/oscillators and show that, although these induce the correct algebraic structures
on  $\mathcal{U}(h)$, they fail to induce the appropriate co-algebraic structures on $\mathcal{U}(h)$. To circumvent this problem, we begin by providing a
brief review of the concept of Wigner oscillators and the  $osp(1|2n)$ super-algebras in section \textbf{6}. With an appropriate interpretation through the
super-algebra structure of the oscillator algebra, we show in section \textbf{7}, that the correct co-algebra structure at the level of $\mathcal{U}(h)$ is also
correctly induced. We then study the physical implications of all these aspects in the context of quantum statistics in section \textbf{8}. The conclusions are contained 
in section \textbf{9}.

\section{Brief Review on Hopf Algebras}
Here we provide a brief review of Hopf algebra and collect some of the essential formulae to be used subsequently. For
this we essentially follow \cite{aschieri, aschieri2}.

If we have a finite Lie algebra $\mathfrak{g}$ comprising
generators $\tau_a$ satisfying

\begin{equation}\label{comm}
[\tau_a,\tau_b]=if_{ab}^c \tau_c
\end{equation} and having the costructures, i.e. the co-product, co-unit and the antipode, given respectively as
($g\in\mathfrak{g}$):
\begin{eqnarray}
   \Delta(g)&=&g\otimes\mathbf{1}+\mathbf{1}\otimes g \\
  \varepsilon(g)&=&0 \\
   S(g)&=&-g,
\end{eqnarray}
we can construct its universal enveloping algebra
$\mathcal{U}(\mathfrak{g})$ which, by definition, contains the
identity $\mathbf{1}$ and polynomials of the generators $\tau_a$
modulo the commutation relations (\ref{comm}). Its algebraic
structures are given by a linear map $\mu$
\begin{eqnarray}
  \mu:\mathcal{U}(\mathfrak{g})\otimes\mathcal{U}(\mathfrak{g})&\rightarrow&\mathcal{U}(\mathfrak{g})\label{mu}\nonumber \\
  \mu(a\otimes b)&=& a\cdot b
\end{eqnarray}
so that $a\cdot\mathbf{1}=\mathbf{1}\cdot a=a$ $ \forall
a\in\mathcal{U}(\mathfrak{g})$, whereas its  costructures are
determined by the following homomorphisms,

\begin{eqnarray}
   \Delta:\mathcal{U}(\mathfrak{g})&\rightarrow&\mathcal{U}(\mathfrak{g})\otimes\mathcal{U}(\mathfrak{g}) \\
  \varepsilon:\mathcal{U}(\mathfrak{g})&\rightarrow&\mathbb{C} \\
   S:\mathcal{U}(\mathfrak{g})&\rightarrow&\mathcal{U}(\mathfrak{g})
\end{eqnarray}
which satisfy linearity for $\Delta$ and $\varepsilon$ and
anti-multiplicativity for $S$
($\xi,\zeta\in\mathcal{U}(\mathfrak{g})$):
\begin{eqnarray}
   \Delta(\xi\zeta)&=&\Delta(\xi)\Delta(\zeta) \label{f1}\\
  \varepsilon(\xi\zeta)&=&\varepsilon(\xi)\varepsilon(\zeta) \\
   S(\xi\zeta)&=&S(\zeta)S(\xi)\label{f3}.
\end{eqnarray}
Additionally, they satisfy
\begin{eqnarray}
   (\Delta\otimes id)\Delta(\xi)&=&(id\otimes \Delta)\Delta(\xi) \text{ (co-associativity)}\\
   (\varepsilon\otimes id)\Delta(\xi)&=&(id\otimes\varepsilon)\Delta(\xi)=\xi \\
   \mu(S\otimes id)\Delta(\xi)&=&\mu(id\otimes
   S)\Delta(\xi)=\varepsilon(\xi)\mathbf{1},
\end{eqnarray}
whereas for the identity $\mathbf{1}\in\mathcal{U}(\mathfrak{g})$
one defines
\begin{eqnarray}
   \Delta(\mathbf{1})&=&\mathbf{1}\otimes\mathbf{1}\\
  \varepsilon(\mathbf{1})&=&\mathbf{1} \\
   S(\mathbf{1})&=&\mathbf{1}.
\end{eqnarray}

With this, the universal enveloping algebra has the structure of a
Hopf algebra.

The Sweedler notation shall be used (a sum over $\xi_1$ and
$\xi_2$ is understood):

\begin{equation}
    \Delta(\xi)=\sum_i \xi_1^i\otimes\xi_2^i\equiv\xi_1\otimes\xi_2
\end{equation}

Now we can deform the Hopf algebra $\mathcal{U}(\mathfrak{g})$
into the Hopf algebra $\mathcal{U}^\mathcal{F}(\mathfrak{g})$ by
means of a twist
$\mathcal{F}\in\mathcal{U}(\mathfrak{g})\otimes\mathcal{U}(\mathfrak{g})$
that is invertible and satisfies the cocycle condition

\begin{equation}\label{cocycle}
    (\mathcal{F}\otimes\mathbf{1})(\Delta\otimes
    id)\mathcal{F}=(\mathbf{1}\otimes\mathcal{F})(id\otimes\Delta)\mathcal{F}.
\end{equation}

The costructures will be deformed as follows

\begin{eqnarray}
   \Delta^\mathcal{F}(g)&=&\mathcal{F}\Delta(g)\mathcal{F}^{-1}\label{defco}\\
  \varepsilon^\mathcal{F}(g)&=&\varepsilon(g) \\
   S^\mathcal{F}(g)&=&\chi S(g)\chi^{-1},
\end{eqnarray}
where
\begin{eqnarray} \chi=f^\alpha S(f_\alpha) \in \mathcal{U}(\mathfrak{g}).\label{chi1} \end {eqnarray}
We are denoting
$\mathcal{F}=f^\alpha \otimes f_\alpha$ and
$\mathcal{F}^{-1}=\bar{f}^\alpha \otimes \bar{f}_\alpha.$ As an algebra $\mathcal{U}^\mathcal{F}(\mathfrak{g})$ is identical
to $\mathcal{U}(\mathfrak{g})$ \cite{aschieri}.

The generators of  $\mathcal{U}^\mathcal{F}(\mathfrak{g})$ are
given as

\begin{equation}
    g\mathcal{^F}=\bar{f}^\alpha(g) \bar{f}_\alpha,
\end{equation}
while their coproduct is
\begin{equation}
 \Delta^\mathcal{F}(g\mathcal{^F})=g\mathcal{^F}\otimes
\mathbf{1}+\bar{R}^\alpha\otimes\bar{R}_\alpha(g\mathcal{^F}),\label{dco}
\end{equation}
with $ \bar{R}^\alpha\otimes\bar{R}_\alpha=\mathcal{R}^{-1}$, where
$\mathcal{R}=(f_\alpha\otimes f^\alpha)\mathcal{F}^{-1}$ is the
universal $\mathcal{R}$-matrix.

Note that it is no longer co-commutative in general. These
generators $g^\mathcal{F}$ form a linear subspace
$\mathfrak{g}^{\mathcal{F}}$ of
$\mathcal{U}^\mathcal{F}(\mathfrak{g})$, the counterpart of
$\mathfrak{g}\subset\mathcal{U}(\mathfrak{g})$.

Finally, the deformed brackets in
$\mathcal{U}^\mathcal{F}(\mathfrak{g})$ are

\begin{equation}\label{defbracket}
    [\xi^\mathcal{F},\zeta^\mathcal{F}]_\mathcal{F}
    =(\xi^\mathcal{F})_1\zeta^\mathcal{F}S(\xi^\mathcal{F})_2,
\end{equation}
where, of course, $(\xi^\mathcal{F})_1
\otimes(\xi^\mathcal{F})_2=\Delta^\mathcal{F}(\xi^\mathcal{F})$
(Sweedler notation).

The deformed brackets satisfy the Jacobi identity

\begin{equation}
    [f,[f',f'']_\mathcal{F}]_\mathcal{F}+
    [f'',[f,f']_\mathcal{F}]_\mathcal{F}+[f',[f'',f]_\mathcal{F}]_\mathcal{F}=0,
\end{equation}
for all $f,f',f''\in \mathfrak{g}^\mathcal{F}$.

The multiplication map $m$ in the module \footnote{This
multiplication map $m$ should not be confused with the linear map
$\mu$ introduced earlier (\ref{mu}). The former acts on a module,
which is just an algebra under $m$, and furnishes a representation
of $\mathfrak{g}$ or, for that matter, of
$\mathcal{U}(\mathfrak{g})$ \cite{aschieri}.} also gets deformed as

\begin{equation}
    m^\mathcal{F}=m \circ\mathcal{F}^{-1}
\end{equation}
to maintain its compatibility with the deformed coproduct
$\Delta^\mathcal{F}(g\mathcal{^F})$ (\ref{defco}).  On the other
hand, the multiplication map $\mu$ undergoes no deformation,
because as an algebra $\mathcal{U}^\mathcal{F}(\mathfrak{g})$ is
the same as $\mathcal{U}(\mathfrak{g})$, as mentioned earlier.

\section{Hopf algebra structure of the (twisted) bosonic Heisenberg algebra}

In the previous section we have provided a general outline of the
Hopf algebra $\mathcal{U}(\mathfrak{g})$ and its deformed
counterpart $\mathcal{U}^{\mathcal{F}}(\mathfrak{g})$. We now
apply it to quantum Heisenberg algebra, involving commutators of
bosonic variables only.

We start with the algebra
$h(\mathcal{N}),\;i,j=1,...,\mathcal{N}$ (in the following we also
consider the limit $\mathcal{N}\rightarrow\infty$). We have

\begin{eqnarray}\label{algebra1}
   \left[x_i,x_j\right]&=&\left[p_i,p_j\right]=0 \label{ha} \\
  \left[x_i,p_j\right]&=&i\delta_{ij}\hat{\hbar} \\
   \left[\hat{\hbar},x_i\right]&=&\left[\hat{\hbar},p_j\right]=0\label{algebra2}
\end{eqnarray}

and apply the twist

\begin{equation}\label{twist}
    \mathcal{F}=\exp\left({\frac{i}{2}\frac{\theta_{ij}}{\hbar^2}p_i\otimes
    p_j}\right),
\end{equation}
where $\theta_{ij}$ is a skew-symmetric matrix. Note that here we do not consider any spacetime
noncommutativity  and therefore set $\theta_{0i}=0$. As the twist
involves only commuting momentum generators, it trivially
satisfies the cocycle condition (\ref{cocycle}).

Before proceeding further, we have to remember that the central
extension of a Lie algebra has to be treated at par with other
generators of the Lie algebra and not as a multiple of the
identity, as  done in \cite{palev}. This is because the
identity belongs to the universal enveloping algebra, whereas the
central extension belongs to the Lie algebra.

In our situation  it is therefore important to note that, although
they represent the same constant,  $\hbar$ in the denominator of
the twist element (\ref{twist}) is a c-number and is introduced
for dimensional reasons, while a hat has been put on $\hat{\hbar}$
occurring in the Lie algebra (\ref{algebra1}-\ref{algebra2}) to
make it explicit that the latter plays the role of a central
extension, having the coproduct
$\Delta(\hat{\hbar})=\hat{\hbar}\otimes
\mathbf{1}+\mathbf{1}\otimes \hat{\hbar}$ and the antipode
$S(\hat{\hbar})=-\hat{\hbar}$.

By using the Baker-Campbell-Hausdorff formula, the deformed
coproduct of $x_k$ can be calculated, by using (\ref{defco}), as

\begin{eqnarray}
    \Delta^\mathcal{F}(x_k)&=&\exp\left({\frac{i}{2}\frac{\theta_{ij}}{\hbar^2}p_i\otimes
    p_j}\right)\Delta(x_k)\exp\left(-{\frac{i}{2}\frac{\theta_{ij}}{\hbar^2}p_i\otimes
    p_j}\right) =\nonumber \\
    &=&x_k\otimes\mathbf{1}+\mathbf{1}\otimes
    x_k+\frac{\theta_{kj}}{2\hbar^2}\left(\hat{\hbar}\otimes p_j-p_j\otimes\hat{\hbar}
    \right).
\end{eqnarray}
Since $p_i$'s and $\hat{\hbar}$ commute with the $p_j$'s, their
coproducts go undeformed:
\begin{eqnarray}
    \Delta^\mathcal{F}(p_k)&=&\Delta(p_k) \\
\Delta^\mathcal{F}(\hat{\hbar})&=&\Delta(\hat{\hbar}).
\end{eqnarray}

Now let us show that the antipode does not get deformed. It
amounts to computing the element (\ref{chi1})
\begin{equation}
    \chi\equiv f^{\alpha}S(f_\alpha)=\exp\left({-\frac{i}{2}\frac{\theta_{ij}}{\hbar^2}p_i
    p_j}\right)=\mathbf{1},
\end{equation}
so that $S^\mathcal{F}=\chi S\chi^{-1}=S.$

Now, the deformed $x_k$ is
\begin{equation}\label{def}
    x_k^\mathcal{F}=\bar{f}^\alpha(x_k)\bar{f}_\alpha=x_k-\frac{1}{2\hbar^2}\theta_{kj}\hat{\hbar}
    p_j,
\end{equation}
while $p_i$ and $\hat{\hbar}$ do not undergo deformation.

In this case, the universal $\mathcal{R}$-matrix is just
$\mathcal{R}=\mathcal{F}^{-2}$, so that the deformed coproduct of
$x_k^\mathcal{F}$ is obtained by using (\ref{dco}) as
\begin{equation}
    \Delta^\mathcal{F}(x_k^\mathcal{F})=x_k^\mathcal{F}\otimes\mathbf{1}+\mathbf{1}\otimes
    x_k^\mathcal{F}+\frac{1}{\hbar^2}\theta_{ik}p_i\otimes \hat{\hbar}.
\end{equation}
It is important to note that the central charge $\hat{\hbar}$
gives a vital contribution to the coproduct.

The antipode of $x_k^\mathcal{F}$ is obtained  by using the anti-multiplicative property of the antipode 
\begin{equation}
    S(x_k^\mathcal{F})=-x_k-\frac{1}{2\hbar^2}\theta_{kj}\hat{\hbar}
    p_j=-x_k^\mathcal{F}-\frac{1}{\hbar^2}\theta_{kj}\hat{\hbar}
    p_j.
\end{equation}

With these expressions in hand, we can calculate the deformed
brackets using (\ref{defbracket})

\begin{eqnarray}
  \left[x_i^\mathcal{F},p_j^\mathcal{F}\right]_\mathcal{F}&=&i\delta_{ij}\hat{\hbar}\\
  \left[x_i^\mathcal{F},x_j^\mathcal{F}\right]_\mathcal{F}&=&0 \\
  \left[p_i^\mathcal{F},p_j^\mathcal{F}\right]_\mathcal{F}&=&0 \\
  \left[\hat{\hbar}^\mathcal{F},x_i^\mathcal{F}\right]_\mathcal{F}&=&\left[\hat{\hbar}^\mathcal{F},p_i^\mathcal{F}\right]_\mathcal{F}=0.
\end{eqnarray}

Note that the deformed brackets of the deformed quantities have the
same structure constants as the undeformed brackets of the
undeformed quantities. The same feature was observed in the case
of the deformed universal enveloping algebra
$\mathcal{U}(iso(1,3))$ of the Poincar\'e algebra \cite{aschieri}.

We can calculate the ordinary brackets of the deformed
quantities:
\begin{eqnarray}
  \left[x_i^\mathcal{F},p_j^\mathcal{F}\right]&=&i\delta_{ij}\hat{\hbar}\\
  \left[x_i^\mathcal{F},x_j^\mathcal{F}\right]&=&\frac{i}{\hbar^2}\theta_{ij}\hat{\hbar}^2 =i\theta_{ij} \label{oc} \\
  \left[p_i^\mathcal{F},p_j^\mathcal{F}\right]&=&0 \\
  \left[\hat{\hbar}^\mathcal{F},x_i^\mathcal{F}\right]&=&\left[\hat{\hbar}^\mathcal{F},p_i^\mathcal{F}\right]=0.
\end{eqnarray}

At this stage we observe that the deformed $x_i^\mathcal{F}$'s
become noncommutative in nature, while the original $x_i$'s were
commutative (\ref{ha}). The inverse transformation of (\ref{def}),
$x_i=x_i^\mathcal{F}+\frac{\theta_{ij}}{2\hbar}p_j$, is an algebra
morphism and is known as Bopp shift in the literature
\cite{bopp}\footnote{In the context of non-linear integrable
systems its analog is known as a dressing transformation
\cite{fadeev}.}. The $x_i$'s have also been identified as the ``classical''
commuting coordinates which are obtained by taking ``average'' of
left and right action of non-commuting $x_i^\mathcal{F}$'s
\cite{bal}. They have also been identified as ``dipole
coordinates'', as they represent certain non-local position
operators which grow with increasing center-of-mass momentum
transverse to their extension, due to their dipole momentum
(see for example, \cite{szabo}). Interestingly, here we find that they have another deep
mathematical attribute viz. they represent the linear subspaces of
$\mathcal{U}(h(\mathcal{N}))$ and
$\mathcal{U}^{\mathcal{F}}({h(\mathcal{N})})$.

In this context let us mention that the $SO(D)$ vectors $x_i$ and
$p_i$ transform as

\begin{eqnarray}
  x_i\rightarrow x_i' &=& U(R)x_iU(R)^\dag=R_{ij}x_j \\
  p_i\rightarrow p_i' &=& U(R)p_iU(R)^\dag=R_{ij}p_j.
\end{eqnarray}
$U(R)$ represents some unitary transformation in an appropriate Hilbert space, corresponding
to the rotation $R\in SO(D)$, which induces the following transformation
on $x_i^\mathcal{F}$:

\begin{equation}
    x_i^\mathcal{F}\rightarrow x_i^{\mathcal{F}'} =
    R_{ij}x_j^\mathcal{F}+\frac{1}{2\hbar}[R,\theta]_{ij}p_j.
\end{equation}

The presence of the $\theta$-dependent inhomogeneous term, which
vanishes only for $D=2$, indicates that $x_i^\mathcal{F}$ does not
transform as a vector under $SO(D)$ for $D>2$. Nevertheless, it
can be easily checked that the commutators (\ref{oc}) transform
as scalar:

\begin{equation}
    [x_i^\mathcal{F},x_j^\mathcal{F}]
\rightarrow[x_i^{\mathcal{F}'},x_j^{\mathcal{F}'}]
=[x_i^\mathcal{F},x_j^\mathcal{F}]=i\theta_{ij},
\end{equation}
as required by the constancy of $\theta_{ij}$.\footnote{One of us, B.C., thanks Sachindeo Vaidya for pointing
this out  to him.} They are regarded
as elements of a constant matrix and not as components of a
transforming tensor, as if representing some new constants of
Nature like $\hbar$, $c$, $G$, etc. \cite{aschieri}.

Finally, the deformed multiplication on the module is

\begin{eqnarray}
    a\star b&\equiv& m^\mathcal{F}(a\otimes b)=(m\circ\mathcal{F}^{-1})(a\otimes
    b)=\bar{f}^\alpha(a)\bar{f}_\alpha(b)=\nonumber \\
   &=& \sum_{n=0}^\infty\frac{1}{n!}\left(\frac{-i}{2\hbar^2}
\right)^n\theta_{i_1j_1}...\theta_{i_nj_n}\left[p_{i_1},...[p_{i_n},a]\right]
\left[p_{j_1},...[p_{j_n},b]\right].
\end{eqnarray}
The module is the space of functions where $x_i$ acts by ordinary
multiplication and $p_i$ acts by differentiation.

Now defining the Moyal bracket, $\left[a,b\right]_\star\equiv
(a\star b - b\star a$), we have
\begin{eqnarray}
  \left[x_i,p_j\right]_\star&=&i\delta_{ij}\hat{\hbar}\\
  \left[x_i,x_j\right]_\star&=&\frac{i}{\hbar^2}\theta_{ij}\hat{\hbar}^2 \\
  \left[p_i,p_j\right]_\star&=&0 \\
  \left[\hat{\hbar},x_i\right]_\star&=&\left[\hat{\hbar},p_i\right]_\star=0.
\end{eqnarray}

This shows that the undeformed brackets of the deformed quantities
have the same structure constants as the Moyal-brackets of the
undeformed quantities.

Having studied the bosonic Heisenberg algebra, we now take up the
fermionic Heisenberg algebra in the next section.

\section{Hopf algebra structure of the (twisted) fermionic Heisenberg algebra}

Let us start with the fermionic algebra
$h_F(\mathcal{N}),\;\alpha,\beta=1,...,\mathcal{N}$.

\begin{eqnarray}
  \left\{\theta_\alpha,\theta_\beta\right\}&=&\left\{\partial_\alpha,\partial_\beta\right\}=0 \\
  \left\{\partial_\alpha,\theta_\beta\right\}&=&\delta_{\alpha\beta}c\\
  \left[c,\partial_\alpha\right]&=&\left[c,\theta_\alpha\right]=0.
\end{eqnarray}
$\theta_\alpha$ and $\partial_\alpha$ are odd generators, while
the central charge $c$ is an even generator.

Analogously, $c$ is the central extension and is treated at par
with the other generators $\theta_\alpha$ and $\partial_\alpha$,
so its coproduct is $\Delta(c)=c\otimes
\mathbf{1}+\mathbf{1}\otimes c$ and the antipode is $S(c)=-c$.

The graded version of the formulae (\ref{f1}-\ref{f3}) is

\begin{eqnarray}
  (\xi\otimes\zeta)(\xi'\otimes\zeta')&=&(-1)^{|\zeta||\xi'|}(\xi\xi'\otimes\zeta\zeta') \\
  S(\xi\zeta)&=&(-1)^{|\xi||\zeta|}S(\zeta)S(\xi)
  \end{eqnarray}
where $|\xi|$ is the degree of $\xi$.

Introducing the twist, in terms of the constant symmetric matrix $C_{\alpha\beta}$,
\begin{equation}
    \mathcal{F}=\exp\left(C_{\alpha\beta}\partial_\alpha\otimes\partial_\beta \right)
\end{equation}
and the graded expressions
  \begin{eqnarray}
  \theta_\beta^{\mathcal{F}}&=&\sum_\alpha(-1)^{|\bar{f}_\alpha||\theta_\beta|}\bar{f}^\alpha(\theta_\beta)\bar{f}_\alpha \\
  \mathcal{R}&=&\sum_{\alpha,\beta}(-1)^{|\bar{f}^\beta||f^\alpha|}(f_\alpha\bar{f}^\beta\otimes f^\alpha\bar{f}_\beta)\\
   \left[u^\mathcal{F},v^\mathcal{F}\right\}_\mathcal{F}&=&\sum_k(u^\mathcal{F})^k_1v^\mathcal{F}(-1)^{|v^\mathcal{F}||(u^\mathcal{F})^k_2|}S(u^\mathcal{F})^k_2,
\end{eqnarray}
one can calculate the deformed coproduct of $\theta_\alpha$:

\begin{equation}
    \Delta^\mathcal{F}(\theta_\alpha)=\Delta(\theta_\alpha)+C_{\alpha\beta}(\partial_\beta\otimes c -
    c\otimes\partial_\beta),
\end{equation}
the others undergoing no deformation.

Since

\begin{equation}
    \chi=f^\alpha
    S(f_\alpha)=\exp\left(C_{\alpha\beta}\partial_\alpha\partial_\beta\right)=\mathbf{1},
\end{equation}
the antipode is also undeformed.

It can be easily seen that only the generator $\theta_\alpha$ gets deformed as

\begin{equation} \label{defteta}
    \theta_\alpha^\mathcal{F}=\theta_\alpha+C_{\alpha\beta}\partial_\beta
    c.
\end{equation}

Its antipode is
\begin{equation}
    S(\theta_\beta^\mathcal{F})=-\theta_\alpha+C_{\alpha\beta}\partial_\beta
    c=-\theta_\alpha^\mathcal{F}+2C_{\alpha\beta}\partial_\beta
    c.
\end{equation}

The universal $\mathcal{R}$-matrix is simply $\mathcal{F}^{-2}$,
so that
\begin{equation}
    \Delta^\mathcal{F}(\theta_\alpha^\mathcal{F})=\theta_\alpha^\mathcal{F}\otimes\mathbf{1}+\mathbf{1}\otimes\theta_\alpha^\mathcal{F}+2C_{\alpha\beta}\partial_\beta\otimes
    c.
\end{equation}

The deformed brackets are now

\begin{eqnarray}
  \left\{\theta_\alpha^\mathcal{F},\partial_\beta^\mathcal{F}\right\}_\mathcal{F}
  &=&\delta_{\alpha\beta}c^\mathcal{F}\\
  \left\{\theta_\alpha^\mathcal{F},\theta_\beta^\mathcal{F}\right\}_\mathcal{F}&=&0 \\
  \left\{\partial_\alpha^\mathcal{F},\partial_\beta^\mathcal{F}\right\}_\mathcal{F}&=&0 \\
  \left[\partial_\alpha^\mathcal{F},c^\mathcal{F}\right]_\mathcal{F}
  &=&\left[\theta_\alpha^\mathcal{F},c^\mathcal{F}\right]_\mathcal{F}=0.
\end{eqnarray}

The ordinary brackets of the deformed quantities are
\begin{eqnarray}
  \left\{\theta_\alpha^\mathcal{F},\partial_\beta^\mathcal{F}\right\}&=&\delta_{\alpha\beta}c \label{d1}\\
  \left\{\theta_\alpha^\mathcal{F},\theta_\beta^\mathcal{F}\right\}&=&2C_{\alpha\beta}c^2 \\
  \left\{\partial_\alpha^\mathcal{F},\partial_\beta^\mathcal{F}\right\}&=&0 \\
  \left[c^\mathcal{F},\theta_\alpha^\mathcal{F}\right]
  &=&\left[c^\mathcal{F},\partial_\alpha^\mathcal{F}\right]=0.
  \label{d4}
\end{eqnarray}

It should be observed once more that the ordinary brackets of the deformed quantities give rise to the above non-anticommutative structure.
This is just analogous to what happens in the bosonic case.

The inverse of (\ref{defteta}), connecting non-anticommutative
variables $\theta_\alpha^\mathcal{F}$ with the anticommuting
$\theta_\alpha$:
$\theta_\alpha=\theta_\alpha^\mathcal{F}-C_{\alpha\beta}\partial_\beta c$,
is the fermionic counterpart of the Bopp shift.

Finally, the deformed multiplication is

\begin{eqnarray}
    a\star b\equiv m^\mathcal{F}(a\otimes b)=(m\circ\mathcal{F}^{-1})(a\otimes
    b)=\sum_\alpha(-1)^{|\bar{f}_\alpha||a|}\bar{f}^\alpha(a)\bar{f}_\alpha(b).
\end{eqnarray}

Additionally, defining $\left[a,b\right\}_\star\equiv a\star b
+(-1)^{|a||b|} b\star a$, we have
\begin{eqnarray}
  \left\{ \theta_\alpha , \theta_\beta\right\}_\star&=&2C_{\alpha\beta}c^2\\
  \left\{\partial_\alpha , \theta_\beta\right\}_\star&=&\delta_{\alpha\beta}c \\
  \left\{\partial_\alpha , \partial_\beta\right\}_\star&=&0 \\
  \left[c,\theta_\alpha\right]_\star&=&\left[c,\partial_\alpha\right]_\star=0.
\end{eqnarray}

Clearly, as in the bosonic case, the fermionic Moyal-brackets are
isomorphic to the ordinary brackets of the deformed quantities
(\ref{d1}-\ref{d4}).

\section{On second quantized operators}

Having studied the Hopf algebra and the deformed Hopf algebra arising from the universal
enveloping algebra of bosonic and fermionic algebras, here we
demonstrate how this Hopf algebra structure can be induced from
the more fundamental Hopf algebra structure of second-quantized
(Schr\"odinger) field-operators or the basic oscillators. This is
important, since elements of the Heisenberg algebra can be expressed in
terms of certain integrated objects of bilinears of those field
operators.

To this end, consider the Schr\"odinger action

\begin{equation}
    S=\int dt L,
\end{equation}
where
\begin{equation}
    L=\int d^Dx \left(\frac{i\hbar}{2}\psi^\ast\stackrel{\leftrightarrow}{\partial_o}\psi-\frac{\hbar^2}{2m}|\vec{\nabla}\psi|^2 \right)
\end{equation}
is the Lagrangian in $D$-dimensional space. It can be easily
checked that it yields the Schr\"odinger equation
\begin{equation}
    i\hbar\frac{\partial \psi}{\partial
    t}=-\frac{\hbar^2}{2m}\nabla^2\psi
\end{equation}
as the Euler-Lagrange equation of motion. One can also check that
the system is subject to the pair of second-class constraints
$\chi\approx 0$ and $\chi^\ast\approx 0$, where $\chi$ is given by
\begin{equation}\label{chi}
\chi=\pi_\psi-\frac{i\hbar}{2}\psi^\ast
\end{equation}
and $\pi_\psi$ represents the canonically conjugate momentum to
$\psi$.

Strong imposition of this pair of constraints results in the
following Dirac brackets \cite{dirac}
\begin{equation}
    \{\psi(\vec{x},t),\psi^\ast(\vec{y},t)\}_{DB}=\frac{1}{i\hbar}\delta^D(\vec{x}-\vec{y}).
\end{equation}

They can be obtained more simply by using the Faddeev-Jackiw
symplectic technique \cite{jackiw}, as already in
first-order form.

They can now be upgraded to the level of quantum commutators for
bosonic systems. Anticommutators should be used for fermions.

\begin{eqnarray}
  \left[ \psi(\vec{x}),\psi^\dag(\vec{y})\right] &=& \delta^D(\vec{x}-\vec{y}) \label{psi1}\\
  \left[ \psi(\vec{x}),\psi(\vec{y})\right] &=& \left[
  \psi^\dag(\vec{x}),\psi^\dag(\vec{y})\right]=0 \label{psi2}
\end{eqnarray}

We now define the following integrated objects involving bilinears
in fields:

\begin{eqnarray}
  X_i &=& \int d^Dy\; y_i \psi^\dag(\vec{y})\psi(\vec{y}) \label{xi}\\
  P_i &=& -\frac{i\hbar}{2} \int d^Dy \;
  \psi^\dag(\vec{y})\stackrel{\leftrightarrow}{\partial_i}\psi(\vec{y})
  \label{P}
\end{eqnarray}
This expression of momentum is obtained from Noether's theorem and by 
making use of the pair of (by now) strong constraints $\chi$
(\ref{chi}) and $\chi^\ast$.

It should be mentioned at this stage that these integrated objects
involving field bilinears provide a mapping from
second-quantization formalism to first-quantization formalism.

Indeed, it can be easily seen, by using (\ref{psi1}, \ref{psi2}),
that $X_i$ can really be identified with the position operator
\begin{equation}
    X_i|\vec{y}\rangle=y_i|\vec{y}\rangle,
\end{equation}
where $|\vec{y}\rangle=\hat{\psi}^\dag(\vec{y})|0\rangle$.

Furthermore, it can now be shown that

\begin{equation}\label{xp}
\left[ X_i,P_j\right]=i\hbar\delta_{ij}N,
\end{equation}

where $N=\int d^Dy\; \psi^\dag(\vec{y})\psi(\vec{y})$ is the
number operator. The rest of commutators vanish, so that $\hbar N$
is identified as the central charge in the $N$-particle sector. One
can also see that

\begin{eqnarray}
  \left[ P_i,\psi(\vec{x})\right]&=& i\hbar\partial_i\psi(\vec{x}) \\
  \left[ P_i,\psi^\dag(\vec{x})\right]&=&
  i\hbar\partial_i\psi^\dag(\vec{x}),
\end{eqnarray}
thus $P_i$ generates the appropriate translations.

It can be recalled that in second quantization the fields are the
operators acting on an appropriate Fock space, which is nothing
but the infinite direct sum of all possible Hilbert spaces
containing all possible number of particles, i.e.,
$\mathcal{H}=\mathcal{H}^{(1)}\oplus\mathcal{H}^{(2)}\oplus...\oplus\mathcal{H}^{(n)}\oplus...$,
where $\mathcal{H}^{(n)}$ is the n-particle Hilbert space and the
coordinate variables play the role of `labels' for the infinite
number of degrees of freedom. On the other hand, in the
first-quantization formalism, the coordinate variables are
themselves the operators acting on an appropriate Hilbert space
and satisfy the Heisenberg algebra along with the conjugate momenta.

It turns out that in non-relativistic quantum mechanics these two
formalisms are completely equivalent, as a generic $N$-particle
state $|\psi_N\rangle$ can be obtained by superposing states, in
terms of first-quantized $N$-particle wavefunctions
$\Phi(\vec{x}_1,...,\vec{x}_N)$, obtained by $N$-fold action of
the creation operators $\hat{\psi}^\dag(\vec{x})$ on the vacuum
state $|0\rangle$ defined as $\hat{\psi}(\vec{x})|0\rangle=0$:
\begin{equation}
    |\psi_N\rangle=\int
    d^Dx_1...d^Dx_N\;\Phi(\vec{x}_1,...,\vec{x}_N)\hat{\psi}^\dag(\vec{x}_1)...\hat{\psi}^\dag(\vec{x}_N)|0\rangle
\end{equation}

 Now the fields can be expanded in their Fourier modes
\begin{eqnarray}
  \psi(\vec{x}) &=& \frac{1}{(2\pi\hbar)^D}\int d^Dp\;e^{\frac{i}{\hbar}\vec{p}\cdot\vec{x}}a_{\vec{p}} \\
  \psi^\dag(\vec{x}) &=& \frac{1}{(2\pi\hbar)^D}\int d^Dp\;e^{-\frac{i}{\hbar}\vec{p}\cdot\vec{x}}a^\dag_{\vec{p}}
\end{eqnarray}

and conversely

\begin{eqnarray}
  a_{\vec{p}} &=& \int d^Dx\;e^{-\frac{i}{\hbar}\vec{p}\cdot\vec{x}}\psi(\vec{x})  \\
  a^\dag_{\vec{p}}&=& \int d^Dx\;e^{\frac{i}{\hbar}\vec{p}\cdot\vec{x}}\psi^\dag(\vec{x}) .
\end{eqnarray}

The algebra of $a_{\vec{p}}$, $a^\dag_{\vec{p}}$  is

\begin{eqnarray}
  \left[ a_{\vec{p}},a^\dag_{\vec{p}'}\right] &=& (2\pi\hbar)^3\delta^D(\vec{p}-\vec{p}') \label{o1} \\
  \left[ a_{\vec{p}},a_{\vec{p}'}\right] &=& \left[ a^\dag_{\vec{p}},a^\dag_{\vec{p}'}\right]=0. \label{o2}
\end{eqnarray}

Now, expressing $\vec{P}$ (\ref{P}) in momentum space as
\begin{equation}
    \vec{P}=\frac{1}{(2\pi\hbar)^3}\int d^Dp\;\vec{p}a^\dag_{\vec{p}}a_{\vec{p}},
\end{equation}
we obtain
\begin{eqnarray}
  \left[ P_i,a_{\vec{p}}\right] &=& -p_ia_{\vec{p}} \\
  \left[ P_i,a^\dag_{\vec{p}}\right] &=& p_ia^\dag_{\vec{p}}.
\end{eqnarray}

One can now construct the Hopf algebra through the universal
enveloping algebra of the centrally extended algebra formed by $a_{\vec{p}},a^\dag_{\vec{p}}$
 \footnote {Note that this algebra is isomorphic to the Heisenberg algebra in the basis of creation
and annihilation operators.}
 and deform it through the same twist element
$\mathcal{F}$ (\ref{twist}), just as before. The linear subspace
of this universal enveloping algebra
$\mathcal{U}^\mathcal{F}(a,a^\dag)$ contains
$a_{\vec{p}}^\mathcal{F}$ and $a^{\dag\mathcal{F}}_{\vec{p}}$, the
deformed version of $a_{\vec{p}}$ and $a^{\dag}_{\vec{p}}$. These
are easily obtained as before to get

\begin{eqnarray}
  a_{\vec{p}}^\mathcal{F} &=& \bar{f}^\alpha(a_{\vec{p}})\bar{f}_\alpha = a_{\vec{p}}\;e^{\frac{i}{2\hbar^2}\theta_{ij}p_iP_j} \label{a1} \\
  a^{\dag\mathcal{F}}_{\vec{p}} &=&\bar{f}^\alpha(a^\dag_{\vec{p}})\bar{f}_\alpha = a^\dag_{\vec{p}}\;e^{-\frac{i}{2\hbar^2}\theta_{ij}p_iP_j}, \label{a2}
\end{eqnarray}

while the deformation of $\psi(\vec{x})$ and $\psi^\dag(\vec{x})$
is:

\begin{eqnarray}
  \psi^\mathcal{F} (\vec{x})&=& \psi(\vec{x})\;e^{\frac{1}{2\hbar}\theta_{ij}\stackrel{\leftarrow}{\partial_i}P_j}  \\
  \psi^{\dag\mathcal{F}}(\vec{x}) &=& \psi^\dag(\vec{x})\;e^{\frac{1}{2\hbar}\theta_{ij}\stackrel{\leftarrow}{\partial_i}P_j}.
\end{eqnarray}

Likewise, they belong to
$\mathcal{U}^\mathcal{F}(\psi,\psi^\dag)$.

It is interesting to note that (\ref{a1},\ref{a2}), relating
deformed and undeformed oscillators, are exactly the same found in
\cite{bal2}.

The deformation of $\psi(\vec{x})$ and $\psi^\dag(\vec{x})$ is
consistent with that of $a_{\vec{p}}$ and $a^\dag_{\vec{p}}$
because

\begin{eqnarray}
  a_{\vec{p}}^\mathcal{F} &=& \int d^Dx\;e^{-\frac{i}{\hbar}\vec{p}\cdot\vec{x}}\psi^\mathcal{F}(\vec{x}) \label{c1}  \\
  a^{\dag\mathcal{F}}_{\vec{p}}&=& \int d^Dx\;e^{\frac{i}{\hbar}\vec{p}\cdot\vec{x}}\psi^{\dag\mathcal{F}}(\vec{x}) \label{c2} .
\end{eqnarray}

The above expressions can be easily proven taking into account that
$\vec{p}\cdot\vec{x}$ is invariant under the twist $\mathcal{F}$:
\begin{equation}
    p_i^\mathcal{F}x_i^\mathcal{F}=p_ix_i=\vec{p}\cdot\vec{x}
\end{equation}

We can therefore identify $\mathcal{U}^\mathcal{F}(a,a^\dag)$ with
$\mathcal{U}^\mathcal{F}(\psi,\psi^\dag)$, as they are basically
obtained by deforming $\mathcal{U}(a,a^\dag)$ and
$\mathcal{U}(\psi,\psi^\dag)$ (the universal enveloping algebra
obtained from $\psi(\vec{x})$ or their Fourier modes). They
therefore correspond to different bases for the primitive
elements.

We can observe at this stage that the algebra satisfied by the
field variables $\psi$ (\ref{psi1}, \ref{psi2}) or their Fourier
amplitudes (\ref{o1}, \ref{o2}) is actually the same as that of the original Heisenberg algebra
$h(N)$ with $N\rightarrow \infty$.
 Indeed, they are infinite in
number, with $\psi^\dag(\vec{x})$ and $a^\dag_{\vec{p}}$ playing
the role of conjugate momentum respectively of $\psi(\vec{x})$ and
$a_{\vec{p}}$, while $\delta^D(\vec{x}-\vec{y})$ and
$(2\pi\hbar)^D\delta^D(\vec{p}-\vec{p}')$ play the role of
central charge instead of $i\hbar\delta_{ij}$. They also have the
nice feature that the deformation induced on one induces a
compatible deformation on the other (\ref{c1},\ref{c2}).

At the level of Lie-algebra, the basic brackets
(\ref{psi1},\ref{psi2}) induce the appropriate brackets of $X_i$ and $P_j$ (\ref{xp}).

Thus, although the algebraic structures of the $\mathcal{N}$-Heisenberg
algebra (\ref{xp}) are obtained from those of $\psi,\psi^\dag$ or
$a, a^\dag$, the costructures are not. To see this, let us try to
naively construct the Hopf algebra structure corresponding to the
universal enveloping algebra of the second-quantized field algebra
(\ref{psi1}, \ref{psi2}) by applying the rules of coproduct
homomorphism (\ref{f1}) valid for bosonic variables. This yields,
using (\ref{xi}),

\begin{eqnarray}
  \nonumber \Delta(X_i)&=&\int d^Dy\; y_i \Delta(\psi^\dag(\vec{y})) \Delta(\psi(\vec{y}))=\\
  \nonumber &=&\int d^Dy\; y_i (\psi^\dag(\vec{y})\otimes\mathbf{1}+\mathbf{1}\otimes\psi^\dag(\vec{y}))(\psi(\vec{y})\otimes\mathbf{1}+\mathbf{1}\otimes\psi(\vec{y}))= \\
  &=& X_i\otimes\mathbf{1}+\mathbf{1}\otimes X_i +\int d^Dy\;
  y_i(\psi^\dag(\vec{y})\otimes\psi(\vec{y})+\psi(\vec{y})\otimes\psi^\dag(\vec{y})).
\end{eqnarray}
Clearly, the presence of the integral involving cross-terms spoils
the expected coproduct of $X_i$, which is expected to be given by
\begin{equation}
    \Delta(X_i)=X_i\otimes\mathbf{1}+\mathbf{1}\otimes X_i.
\end{equation}
We solve this problem in Section {\bf 7}, using the concepts of Wigner oscillators and super-algebras.
We therefore begin by providing a brief review in the next section.

\section{Wigner Oscillators and Superalgebras}

The notion of Wigner's oscillators derives from \cite{wigner} (see also \cite{palev2}, which started a series of related works; updated references can be found in \cite{lievens}). In that work Wigner realized that the Hamiltonian equations can be identical to the Heisenberg equations for position and momentum operators, without necessarily realizing the
canonical commutation relations.  For oscillators, the Wigner's consistency conditions induce non-linear relations,
involving commutators and anticommutators of the position and momentum operators, which can be recast as superalgebras.
\par
The simplest example, for a single bosonic oscillator,  requires the hamiltonian $H$ to be expressed as an anticommutator of the oscillators $a^\pm$.
\footnote{In the following, we shall use the notation $a=a^{-}$ and  $a^{\dagger}=a^{+}$.} By defining
\begin{eqnarray}
H&=& \frac{1}{2}\{a,a^\dagger\},\nonumber\\
\relax [ H,a^\pm ] &=& \pm a^\pm\end{eqnarray}\footnote{This second equation, involving the commutator with $H$,  represents the compatibility condition between the Heisenberg
equation and the operator-valued Hamilton's equation.} and setting
\begin{eqnarray}
E^\pm &=& \{a^\pm,a^\pm\},
\end{eqnarray}
one recovers that $H, a^\pm, E^\pm$ are a set of generators of the $osp(1|2)$ superalgebra (see \cite{fss} for
 a quick introduction to $osp(1|2)$ and the other superalgebras). $H$ is the Cartan element, while $a^\pm$ are the fermionic simple roots and correspond to the odd sector
of the superalgebra. It should be noticed that the superalgebra interpretation requires $a^\pm$ to be odd generators, and therefore being of opposite statistics w.r.t. the usual interpretation of $a^\pm$ as bosonic creation and annihilation operators
satisfying the Heisenberg algebra. The hamiltonian $H$, nevertheless, being bilinear in $a^\pm$, keeps its bosonic
 character. The connection of the hamiltonian (conveniently normalized as $H'=\frac{1}{2}H$) with the ordinary harmonic oscillator hamiltonian (normalized s.t. $\omega\hbar=1$) is made
 in terms of the highest weight representations of $osp(1|2)$. The energy levels of the oscillator hamiltonian are
 given by $E_n =\frac{1}{2}+n$, for non-negative $n$ ($\frac{1}{2}$ is the vacuum energy). On the other hand, the
 highest weight representations of $osp(1|2)$, defined by the more general condition $a^-|0>=0$ and $H'|0>=\lambda |0>$,
 imply that the energy spectrum of $H$ is given by a set of bosonic energy levels, whose eigenvalues are
 $E_n=\lambda +n$, for $(E^+)^n|0>$ eigenvectors, as well as a set of fermionic energy levels given by
 $E_n = \lambda +n+\frac{1}{2}$, for the associated $(a^+)^{2n+1}|0>$ eigenvectors. Disregarding the fermionic sector, the bosonic
 sector reproduces the eigenvalues of the ordinary harmonic oscillator for $\lambda=\frac{1}{2}$.\par
 The Wigner's approach can be extended to several sets of both bosonic and fermionic creation and annihilation operators (originally satisfying the bosonic and fermionic Heisenberg algebras). In this case the Wigner's construction induces
 superalgebras of $osp(m|n)$ and $sl(m|n)$ series (see e.g. \cite{palev3}).\par
 A somehow different connection between superalgebras and bosonic and fermionic oscillators can be found in \cite{tang}.
 There,  several series of superalgebras are realized in terms of an oscillator construction requiring $m$ fermionic oscillators $f_i, f_i^\dagger$ and $n$ bosonic oscillators $a_j, a_j^\dagger$. An explicit construction
was given for the superalgebras of the $A(m-1|n-1)=sl(m|n)$, $B(m|n)=osp(2m+1|2n)$ and $D(m|n)=osp(2m|2n)$ series.
 As an example, the $osp(2m|2n)$ superalgebra whose bosonic sector coincides with $so(2m)\oplus sp(2n)$
and whose total number of generators is $2m^2+2n^2-m+n$ bosonic and $4mn$ fermionic, is reproduced by the whole set of
bilinear combinations of $m$ fermionic and $n$ bosonic oscillators.

\section{Hopf algebra structure of the second-quantized operators}

The second quantization requires the introduction of operators constructed with bilinear
combinations of (bosonic and/or fermionic) creation and annihilation operators.
We are interested in such operators like the oscillator-number $N$, the position $\vec{X}$, the momentum
$\vec{P}$, as well as the functions thereof.  The hamiltonian $H$ is given, e.g., by $H=\frac{1}{2m} P^2 + V(\vec{X})$.
An interesting class of operators is given by the bilinear combinations of $\vec{P}$ and $\vec{X}$, namely
$P^2$, $X^2$ and $\frac{1}{2}(\vec{X}\vec{P}+\vec{P}\vec{X})$, which will be discussed later.
The bosonic ($a_i$, $a^\dag_i$) and fermionic ($b_j$, $b^\dag_j$) creation and annihilation operators satisfy
the bosonic and, respectively, the fermionic Heisenberg algebra introduced in the previous sections. As discussed there,
both the bosonic and the fermionic
Heisenberg algebra must be replaced by a graded Lie algebra. The universal enveloping algebra of
the Heisenberg algebra acquires the status of a Hopf algebra, which can be eventually deformed with a twist.\par
In general, for a finite Lie algebra $g$ of dimension $N$, which admits a central extension $c$ (we denote as $\tau_i$ the set of generators of $g$, while ${\check{\tau}}_j$ denotes the subset of generators which do not coincide with $c$),
we can introduce an ordering in its generators ($\tau_1<\tau_2<\ldots <\tau_N$). The universal enveloping
algebra ${\cal U}(g)$ can be decomposed, as a vector space, into
\begin{eqnarray}
{\cal U}(g)&=& {g}_0\oplus {g}_1\oplus {g}_2\oplus \ldots
\end{eqnarray}
where $g_0$ coincides with the identity,
\begin{eqnarray}
g_0 &=& {\bf 1},
\end{eqnarray}
$g_1$ coincides with the Lie algebra $g$ itself,
\begin{eqnarray}
g_1 &=& g,
\end{eqnarray}
while $g_k$ is spanned by the ordered $k$-ples of the $\tau_i$ generators,
\begin{eqnarray}
\tau_{i_1}\tau_{i_2}\ldots \tau_{i_k} &\in&g_k
\end{eqnarray}
for $\tau_{i_1}\leq \tau_{i_2}\leq \ldots \leq \tau_{i_k}$.\par
The $g_k$ space can be further decomposed into its $h_k^l$ subspaces,
\begin{eqnarray}
g_k&=& {h}_k^0\oplus {h}_k^1\oplus \ldots\oplus \ldots h_k^k,
\end{eqnarray}
s.t. $l$ denotes the power of $c$ entering the decomposition. Therefore
$h_k^0$ is spanned by $k$-ples of ordered ${\check{\tau}}_j$ generators,
while, symbolically, $h_k^l \equiv c^l h_{k-l}^0$.\par
Taking into account that the only generator entering the non-vanishing r.h.s. in the (bosonic and fermionic) Heisenberg
algebras is the central extension $c$, we can therefore conclude that
\begin{eqnarray}
[h_2^0,h_2^0] \subset h_2^1.
\end{eqnarray}
It implies that the bilinear combinations of ${\check{\tau}}_j$ generators acquire a Lie algebra
structure, provided that $c$, the central extension, would be re-interpreted as a c-number.
The Lie algebra structure on $h_2^0$ is naturally induced by the Lie algebra structure on $g$. On the other
hand, as a Lie algebra, $h_2^0$ induces a Hopf algebra structure in its universal enveloping algebra
${\cal U}(h_2^0)$. This Hopf algebra structure is not related with the original Hopf algebra structure
defined on ${\cal U}(g)$ and in particular is not a sub-Hopf algebra of the ${\cal U}(g)$ Hopf algebra.  This
is due to the different role of $c$, entering as a central extension in $g$ and as a c-number in $h_2^0$.\par
The resulting observation is that, while it is still possible to define a Hopf algebra structure for
bilinear second-quantized
operators, their ``composite nature" is lost in the process. Their Lie algebra is determined by a more fundamental
(the ``oscillators") level. Their co-structures however, in particular the coproduct, are not. This situation is clearly unsatisfactory. We have seen for instance, for the class of deformations arising from a twist, that an ``ideological
viewpoint" can be maintained: within a suitable basis the Hopf algebra deformation is carried by the co-structures
alone (in particular, the coproduct). A twist deformation of second-quantized operators would be hand-imposed and
would spoil the connection with oscillators. Conversely, a twist deformation of the oscillators would not reflect in
a twist deformation of second-quantized operators.\par
This situation can be reconciled by making use of the Wigner's approach to the oscillators algebra.
As discussed in Section {\bf 6} the troublesome Heisenberg algebras are replaced by superalgebras which do not admit
central extension. The ordinary oscillators' energy eigenvalues
are recovered as specific highest weight representations. This construction, besides providing the Hopf algebra structure
for the ordinary oscillators, allows their extension and deformations. Their extension, already discussed in
Wigner's original paper \cite{wigner}, refers to the choice of the highest weight representation. The deformation
is induced by such deformations (like the twists, see \cite{borowiec} and \cite{celeghini}) of the original superalgebra
which preserve the graded Hopf algebra structure.\par

All that we have to do at this stage is to re-write the operators $X_i$ (92) and $P_i$ (93) and the number operator $N$
in Weyl-symmetric form:

\begin{eqnarray}
  \tilde{X}_i &=& \frac{1}{2} \int d^Dy\; y_i \left( \psi^\dag(\vec{y})\psi(\vec{y}) + \psi(\vec{y})\psi^\dag(\vec{y})  \right) \label{n1} \\
  \tilde{P}_i &=& \frac{1}{2} \int d^Dp\; p_i \left( a^\dag_{\vec{p}} a_{\vec{p}} + a_{\vec{p}} a^{\dag}_{\vec{p}} \right)\\
  \tilde{N} &=& \frac{1}{2} \int d^Dy\;  \left( \psi^\dag(\vec{y})\psi(\vec{y}) + \psi(\vec{y})\psi^\dag(\vec{y})  \right)\label{n3}
\end{eqnarray}

It can now be easily checked that they satisfy the same algebra as the untilded operators. Note that to facilitate subsequent computations,
we have re-written the expression of $ \tilde{P}_i  $ in momentum space, where it takes the diagonal form. The other two variables
$\tilde{X}_i $ and $\tilde{N} $ have similar forms already in coordinate space itself.

If we now declare $\psi(\vec{y})$ and $a_{\vec{p}}$ to be odd, the coproduct of $\tilde{X}_i$ is correctly induced as

\begin{eqnarray}
  \nonumber \Delta(\tilde{X}_i)&=&\frac{1}{2}\int d^Dy\; y_i (\Delta(\psi^\dag(\vec{y})) \Delta(\psi(\vec{y}))+\Delta(\psi(\vec{y})) \Delta(\psi^\dag(\vec{y})))=\\
  \nonumber &=&\frac{1}{2}\int d^Dy\; y_i [\psi^\dag(\vec{y})\psi(\vec{y})\otimes\mathbf{1}-
  \psi(\vec{y})\otimes\psi^\dag(\vec{y})+
  \psi^\dag(\vec{y})\otimes\psi(\vec{y})+\mathbf{1}\otimes\psi^\dag(\vec{y})\psi(\vec{y})+\\  \nonumber &&+\psi(\vec{y})\psi^\dag(\vec{y})\otimes\mathbf{1}-
  \psi^\dag(\vec{y})\otimes\psi(\vec{y})+\psi(\vec{y})\otimes\psi^\dag(\vec{y})+
  \mathbf{1}\otimes\psi(\vec{y})\psi^\dag(\vec{y})] =\\
  &=& \tilde{X}_i\otimes\mathbf{1}+\mathbf{1}\otimes \tilde{X}_i ,
\end{eqnarray}
the same holding for the coproduct of $\tilde{P}_i$.

The antipode is also properly induced as

\begin{eqnarray}
  \nonumber S(\tilde{X}_i)&=& \frac{1}{2}\int d^Dy\; y_i (S(\psi(\vec{y})\psi^\dag(\vec{y}))S(\psi^\dag(\vec{y})\psi(\vec{y}))=\\
  \nonumber &=& \frac{1}{2}\int d^Dy\; y_i[(-1)^{|\psi(\vec{y})||\psi^\dag(\vec{y})|}S(\psi^\dag(\vec{y}))
  S(\psi(\vec{y}))+(-1)^{|\psi^\dag(\vec{y})||\psi(\vec{y})|}S(\psi(\vec{y}))
  S(\psi^\dag(\vec{y}))] =\\
  &=&\frac{1}{2} \int d^Dy\; y_i \left( -\psi^\dag(\vec{y})\psi(\vec{y}) - \psi(\vec{y})\psi^\dag(\vec{y})\right)=
  \nonumber \\
  &=&-\tilde{X}_i,
\end{eqnarray}
as well as
\begin{equation}
    S(\tilde{P}_i)=-\tilde{P}_i
\end{equation}

Co-unit poses no problem since
$\varepsilon(\psi(\vec{y}))=\varepsilon(a_{\vec{p}})=0$, leading to $\varepsilon(\tilde{X}_i)=\varepsilon(\tilde{P}_i)=0$.

Expressions (\ref{n1}-\ref{n3}) yield the expected (absence of) deformation for $\tilde{P}_i$:

\begin{eqnarray}
  \tilde{P}_i^\mathcal{F} &=& \frac{1}{2} \int d^Dp\; p_i \left( a^{\dag\mathcal{F}}_{\vec{p}} a^\mathcal{F}_{\vec{p}} + a^\mathcal{F}_{\vec{p}} a^{\dag\mathcal{F}}_{\vec{p}} \right)=\nonumber \\
  &=& \frac{1}{2} \int d^Dp\; p_i \left( a^\dag_{\vec{p}}\,e^{\frac{i}{2\hbar^2}\theta_{ij}p_iP_j}e^{\frac{-i}{2\hbar^2}\theta_{ij}p_iP_j} a_{\vec{p}} + a_{\vec{p}}\,e^{\frac{-i}{2\hbar^2}\theta_{ij}p_iP_j}e^{\frac{i}{2\hbar^2}\theta_{ij}p_iP_j} a^{\dag}_{\vec{p}} \right)  =\nonumber\\
  &=& \tilde{P}_i,
\end{eqnarray}

as well as for $\tilde{X}_i$, as can be seen in momentum space:

\begin{equation}\label{xtil}
      \tilde{X}^\mathcal{F}_i = \frac{i\hbar}{4} \int d^Dp \;\left(
  a^{\dag\mathcal{F}}_{\vec{p}}\stackrel{\leftrightarrow}{\partial_{p_i}}a^\mathcal{F}_{\vec{p}}+a^\mathcal{F}_{\vec{p}}\stackrel{\leftrightarrow}{\partial_{p_i}}a^{\dag\mathcal{F}}_{\vec{p}}\right)=\tilde{X}_i-\frac{1}{2\hbar^2}\theta_{ij}p_j\hbar\tilde{N},
\end{equation}
i.e., it reduces to the previously obtained deformation
(\ref{def}) at the one particle ($\mathcal{N}=1$) limit.

\section{Hopf algebra and quantum statistics}

We discuss now the relation between Wigner's oscillators (Wigner's approach and extension of the standard oscillator
algebra), Hopf algebras and quantum statistics. Related issues have been discussed in \cite{crv1} and \cite{crv2}.
However, some comments are necessary.  It is sufficient to discuss the single bosonic oscillator which,
in Wigner's approach, is related  to a given highest weight representation of $osp(1|2)$. The vacuum state
$|0>$ is assumed to be bosonic. Due to the fermionic nature of $a^+$, which belongs to the odd sector of the $osp(1|2)$
superalgebra, applying integer powers ${a^+}^k$ to the vacuum produces a tower of states which are, alternately, bosonic
and fermionic. If $\lambda=\frac{1}{2}$ ($H'|0>=\lambda|0>$) one can introduce the fermion-number operator $N_F$, which in the $osp(1|2)$
Wigner realization of the oscillator algebra, can be expressed as
\begin{eqnarray}
N_F &=& \frac{1}{2}(1+e^{2\pi i H'}).
\end{eqnarray}
The bosonic sector is recovered through the superselection rule $N_F=0$ (the corresponding projector is
${\bf 1}-N_F$). The fermionic sector has eigenvalue $N_f=1$.
The energy eigenvalues $E_n^{bos}$ of the bosonic states are given by $E_n^{bos}= \lambda +n$,
for $n=0,1,2,\ldots$, while the energy eigenvalues $E_n^{fer}$ of the fermionic eigenstates
are $E_n^{fer}= \lambda +\frac{1}{2}n$ for $n=1,2,3,\ldots$. 

 The standard oscillator Hilbert space is therefore recovered from the $osp(1|2)$  $\lambda=\frac{1}{2}$ highest weight representation by taking the bosonic sector. This is
tantamount to start, from the very beginning, by looking at a specific highest weight representation of $su(1|1)$ (the $osp(1|2)$ bosonic subalgebra), whose generators (suitably normalized) are $H,E^{\pm}$.\par
On the other hand, limiting the construction to the $su(1|1)$ bosonic algebra would prevent rexpressing the hamiltonian
in terms of a bilinear combination of oscillators (the anticommutator (up to a normalizing factor) of $a^\pm$).\par
The theory derived from $osp(1|2)$ is more general because it produces both bosonic and fermionic eigenstates of the hamiltonian. Taking the bosonic projection is an extra requirement which needs not to be necessarily implemented.\par
Contrary to what stated in \cite{crv2}, the $osp(1|2)$ spectrum is {\em not} supersymmetric. The reason is due to the fact that there is no degeneracy of
the positive energy eigenvalues between fermionic and bosonic states.
In order to fulfil a true supersymmetry, the presence of at least one bosonic oscillator and one fermionic oscillator
satisfying the Heisenberg (respectively bosonic and fermionic) algebra is required. In the Wigner interpretation, this coupled
system of oscillators has to be replaced by its associated superalgebra. The results of \cite{tang}, recalled in Section {\bf 6}, show that one such superalgebra is $osp(2|2)$. For such superalgebra it is indeed possible to introduce a hamiltonian
operator (bilinear w.r.t. the odd generators of $osp(2|2)$) s.t. its spectrum produces the supersymmetric degeneracy
with a one-to-one correspondence of bosonic and fermionic eigenstates for every positive eigenvalue of the energy.
The detailed construction will be produced elsewhere. Other superalgebras are related with other systems of oscillators; \cite{palev3}, e.g., relates a $3$-dimensional oscillator to the $osp(3|2)$ superalgebra.\par

The interpretation of the coproduct follows now the one given in \cite{crv2}.
Let us focus on the $su(1,1)$ algebra (the generalization to other algebras and superalgebras is straightforward) expressed by the $H, E^\pm$ generators satisfying
\begin{eqnarray}
\relax [H, E^\pm] &=& \pm 2 E^\pm,\nonumber\\
\relax [E^+,E^-] &=& H.
\end{eqnarray} 
In the Wigner's interpretation, the hamiltonian for the harmonic oscillator can be expressed as
${\bf H}= \frac{\omega}{2}H$. 
If one starts with a highest-weight vector $|0\rangle$ s.t. $E^-|0\rangle=0$ and  $H|0\rangle=\mu|0\rangle$, therefore, for $su(1|1)$ the hamiltonian ${\bf H}$ admits eigenvalues $\omega(\frac{\mu}{2}+m)$ when applied to its ${E^+}^m|0\rangle$ eigenvector.
Setting ${\widetilde{ |m\rangle}} = {E^+}^m|0\rangle$, with straightforward computations,
one can introduce the normalized state 
\begin{eqnarray}
|m\rangle &=& \frac{1}{\sqrt{(-1)^m m! \prod_{j=0}^{m-1} (\mu+j)}}{\widetilde {|m\rangle}},
\end{eqnarray}
where
\begin{eqnarray}
\langle m|m\rangle &=& 1.
\end{eqnarray}
 The integer $m$ can be regarded both as an energy level or as an $m$-particle state. Let us call ${\cal H}$ the Hilbert space associated to the highest weight representation of ${\bf H}$. The coproduct
$\Delta^n {({E^+}^m)}$ induces the map ${{E^+}^m|0\rangle}\in {\cal H} \mapsto {\cal H}\otimes \ldots \otimes {\cal H}$ ($\equiv {\cal H}^{\otimes n+1}$). The $j$-th Hilbert space in the tensor product can be referred to as the $j$-th slot.  The hamiltonian acting on the tensor product is $\Delta^n ({\bf H})$. The vacuum state in ${\cal H}^{\otimes n+1}$ is given by the tensor product $|0\rangle \otimes \ldots \otimes |0\rangle\equiv |{\bf 0}\rangle$.
The probability that an $m$-particle state can be realized with $m_j$ particles at the $j$-th slot
(s.t. $\sum_j m_j=m$) depends on the highest weight $\mu$.
\par
We explicitly discuss the $n=1$, $m=2$ example. The (unnormalized) state
$\Delta ({E^+}^2)|{\bf 0}\rangle$ is given by
\begin{eqnarray}
\Delta ({E^+}^2)|{\bf 0}\rangle&=& \sqrt{2\mu(\mu+1)}(|2\rangle\otimes|0\rangle+|0\rangle\otimes|2\rangle)-2\mu|1\rangle\otimes|1\rangle.
\end{eqnarray}
The normalized state is
\begin{eqnarray}
|{\bf 2}\rangle &=& \frac{1}{\sqrt{4\mu(1+2\mu)}}\Delta ({E^+}^2)|{\bf 0}\rangle.
\end{eqnarray}
The probability to recover, e.g.,  one particle in the fist slot and one particle in the second slot is given by
$P_{11} = | (\langle 1|\otimes \langle 1|)|{\bf 2}\rangle|^2$ (and similarly for the other cases $P_{20} = | (\langle 2|\otimes \langle 0|)|{\bf 2}\rangle|^2$ and
$P_{02} = | (\langle 0|\otimes \langle 2|)|{\bf 2}\rangle|^2$). These probabilities are explicitly given by
$P_{11}= \frac{\mu}{2\mu+1}$, $P_{20}=P_{02}= \frac{\mu+1}{4\mu+2}$. The (Bose-Einstein) equipartition, $P_{11}=P_{20}=P_{02}=\frac{1}{3}$ is recovered for the highest weight $\mu=1$.\par
This analysis can be repeated for other algebras and superalgebras, recovering as well the Fermi-Dirac statistics. An important final comment is the following: not only deformations of the algebra (the deformed coproduct) can change the usual framework, but also the choice
of the vacuum energy (specified by the highest weight $\mu$). Statistics can therefore be deformed with two different prescriptions.

\section{Conclusions}
We have shown that it is indeed possible to define a Hopf algebra structure on the universal enveloping algebra $\mathcal{U}(h)$ of the Heisenberg
algebra $h$ and deform it to $\mathcal{U}^{\mathcal F}(h)$
by applying the abelian twist {\it \`a la} Drinfeld, which satisfies the co-cycle condition trivially, provided that the role of the 
central extension is identified properly. To be more precise, we require
 the central extension to be regarded as a generic element of the Lie algebra, enjoying the same co-algebra structure as the other generators
$x_i$ and $p_j$ of the Heisenberg algebra $h$ and not merely as a multiple of the identity  (which by definition is included in
$\mathcal{U}(h)$) of the corresponding Lie group. We have shown that the deformed generators $x_i^{\mathcal F}$ and $p_j^{\mathcal F}$, the primitive elements spanning the linear sub-space
of $\mathcal{U}^{\mathcal F}(h)$ under the deformed commutator $[,]_{\mathcal F}$,  have an isomorphic structure to those of the undeformed version.
In particular, the commuting $[x_i,x_j]=0$ subalgebra maps again into the corresponding commuting version under complete deformation:
$[x_i^{\mathcal F},x_j^{\mathcal F}]_{\cal F}=0$. The non-commutativity is obtained only in the hybrid case, i.e. when the ordinary commutators
of the deformed variables are computed $[x_i^{\mathcal F},x_j^{\mathcal F}]=i\theta_{ij}$. We have also shown how this implies that the deformed
$x_i^{\mathcal F}$'s no longer have a vectorial transformation property under $SO(D)$ rotation (for $D>2$) to conform to constant and
non-transforming $\theta_{ij}$ tensor. The parallel of this analysis involving bosonic variables is also valid for fermionic variables.
\\
The position and momentum generators of the Heisenberg algebra can be regarded as composite objects expressed in terms  of integrated bilinears of the Schr\"odinger fields/oscillators.  The oscillator algebra can be upgraded to its
universal enveloping algebra, with its own Hopf algebra structure, and can be deformed by a twist element. However, this Hopf algebra structure
does not induce the appropriate Hopf algebra structure for the Heisenberg algebra defined by the composite position and momentum generators. The failure is at the level of
co-algebra; indeed, due to certain ambiguous roles played
by the central extension, the correct co-product is not induced. A solution to this problem and a full Hopf algebra mapping can be obtained by getting rid, altogether, of the central charge and by making use of the Weyl ordering. Indeed, using the Wigner's prescription, we can regard the bosonic Schr\"odinger oscillators as odd generators of an appropriate super-algebra, such as $osp(1|2n)$. By correctly taking into account the odd nature of the oscillators in the Wigner's prescription, it is indeed possible to induce
a Hopf algebra structure on $\mathcal{U}(h)$ and  deform it to get $\mathcal{U}^{\mathcal F}(h)$.

We have also discussed the implication of this construction for quantum statistics, showing that both the deformed co-product and the choice of the vacuum
energy corresponding to the highest weight representation can give rise to deformations for ordinary Bose-Einstein and Fermi-Dirac
statistics.
{}~
\\{}~
\par {\large{\bf Acknowledgments}}{} ~\\{}~\par
B. C. acknowledges TWAS-UNESCO associateship appointment at CBPF
and CNPq for financial support  and CBPF,
where this work was initiated, for hospitality. B. C. would also
like to thank S. Vaidya for some useful comments. F. T. is
grateful to P. Aschieri for useful comments on the Hopf algebra
structure of the universal enveloping algebras. P. G. C.
acknowledges financial support from CNPq.


\begin{thebibliography}{99}
\bibitem{szabo1} R.~Szabo, Phys.Rep. {\bf 378} (2003) 207; E.~Akofor, A.~P.~Balachandran and A.~Joseph, 
arXiv:0803.4351 (hep-th).

\bibitem{ban} R.~Banerjee, B.~Chakraborty and K.~Kumar, Phys.Rev.D {\bf 71} (2005) 085005; Phys.Rev.D {\bf 77} (2008) 048702.

\bibitem{scholtz} F.~G.~Scholtz, B.~Chakraborty, S.~Gangopadhyay and A.~Ghosh Hazra, Phys.Rev.D{\bf 71} (2005) 085005.

\bibitem{aschieri1} P.~Aschieri, 
arXiv:hep-th/0608172.

\bibitem{doplicher} S.~Doplicher, K.~Fredenhagen and J.~Roberts, Commun. Math.Phys. {\bf 172} (1995) 187; Phys. Lett. B{\bf 331} (1994) 39.

\bibitem{aschieri} P.~Aschieri, M.~Dimitrijevic, F.~Meyer and J.~Wess,
  Class.\ Quant.\ Grav.\  {\bf 23}, 1883 (2006)
  [arXiv:hep-th/0510059].

\bibitem{chaichian} M.~Chaichian, P.~Kulish, K.~Nishijima and A.~Tureanu, Phys. Lett. B{\bf 604}(2004) 98.

\bibitem{chakraborty} B.~Chakraborty, S.~Gangopadhyay, A.~Ghosh Hazra and F.~G.~Scholtz, J. Phys.A {\bf 39} (2006) 9557.

\bibitem{palev}
  T.~D.~Palev,
  arXiv:hep-th/9307032.

   A.~O. Barut, R.~Raczka.  ``Theory of group representations and applications.''   2. ed.  World Scientific Publishing Co., Singapore
   (1986).

\bibitem{aschieri2} P.~Aschieri, 
arXiv: hep-th/0703013.

\bibitem{bopp} F.~Bopp,
  Ann.\ Inst.\ Henri\ Poincar\'e\ {\bf 15} (1956) 81-112.




  \bibitem{fadeev}
  L.~D.~Faddeev, Sov.\ Sci.\ Rev.\ Math.\ Phys.\ {\bf C1} (1980) 107-155.

  H.~Grosse,
   Phys.\ Lett.\  B {\bf 86} (1979) 267-271.

  A.~B.~Zamolodchikov and Al.~B.~Zamolodchikov,
  Ann.\ Phys.\ {\bf 120} (1979) 253.

\bibitem{bal}
  A.~P.~Balachandran, A.~Pinzul, B.~A.~Qureshi, S.~Vaidya and I.~I.~S.~Bangalore,
  arXiv:hep-th/0608138.



  \bibitem{szabo}
  M.~Riccardi and R.~J.~Szabo,
  arXiv:0711.1525 [hep-th].

  \bibitem{dirac}
  P.~A.~M.~Dirac, ``Lectures on Quantum Mechanics'', Belfer
  Graduate School of Science, Yeshiva University, New York (1964).


\bibitem{jackiw}
  L.~D.~Faddeev and R.~Jackiw,
  Phys.\ Rev.\ Lett.\  {\bf 60}, 1692 (1988).

  \bibitem{bal2}
  A.~P.~Balachandran, G.~Mangano, A.~Pinzul and S.~Vaidya,
  Int.\ J.\ Mod.\ Phys.\  A {\bf 21}, 3111 (2006)
  [arXiv:hep-th/0508002].

\bibitem{wigner}
  E.~P.~Wigner,
   Phys.\ Rev.\  {\bf 77}, 711 (1950).

  \bibitem{palev2}
  T.~D.~Palev,
  J.\ Math.\ Phys.\ {\bf 23}, 1778 (1982)

  \bibitem{lievens}
  S.~Lievens, N.~I.~Stoilova and J.~Van der Jeugt,
  arXiv:0709.0180 [hep-th].


\bibitem{fss}
  L.~Frappat, P.~Sorba and A.~Sciarrino,
  arXiv:hep-th/9607161.

  \bibitem{palev3}
  T.~D.~Palev and N.~I.~Stoilova,
  J.\ Phys.\ A  {\bf 27}, 7387 (1994)
  [arXiv:hep-th/9405125].

\bibitem{tang}
  D.~S.~Tang,
  J.\ Math.\ Phys.\  {\bf 25}, 2966 (1984).

  \bibitem{borowiec}
  A.~Borowiec, J.~Lukierski and V.~N.~Tolstoy,
  Mod.\ Phys.\ Lett.\  A {\bf 18}, 1157 (2003)
  [arXiv:hep-th/0301033].

\bibitem{celeghini}
  E.~Celeghini and P.~P.~Kulish,
  J.\ Phys.\ A  {\bf 37}, L211 (2004)
  [arXiv:math/0401272].

  \bibitem{crv1} E.~Celeghini, M.~Rasetti and G.~Vitiello, J.\ Phys.\ A {\bf 28}, L239
  (1995).

  \bibitem{crv2} E.~Celeghini, M.~Rasetti and G.~Vitiello, J.\ Phys.\ A: Math. Gen. {\bf 30}, L125  (1997).


\end{thebibliography}
\end{document}